\documentclass[amsmath,amssymb,aps,preprintnumbers,twocolumn,nofootinbib]{revtex4-1}
\usepackage{epsf}

\usepackage{epsfig,graphics}
\usepackage {graphicx}
\usepackage {epsfig}
\usepackage {psfig}
\usepackage {subfigure}
\usepackage {tabularx} 
\usepackage{rotate}	
\usepackage{slashed}
\usepackage{float}
\usepackage{bbm}
\usepackage{dsfont}
\usepackage[usenames,dviptsnames]{xcolor}

\usepackage{amsmath}
\usepackage{amsfonts}
\usepackage{amssymb}
\usepackage{graphicx}

\newcommand{\bmat}{\left(\begin{array}}
\newcommand{\emat}{\end{array}\right)}

\def\yzero{\smash{\hbox{$y\kern-4pt\raise1pt\hbox{${}^\circ$}$}}}

\def\beq{\begin{equation}}
\def\eeq{\end{equation}}
\def\beqa{\begin{eqnarray}}
\def\eeqa{\end{eqnarray}}

\def\-{\hphantom{-}}

\def\s2{\frac{1}{\sqrt2}}

\def\beq{\begin{equation}}
\def\eeq{\end{equation}}
\def\beqa{\begin{eqnarray}}
\def\eeqa{\end{eqnarray}}
\def\t^{\rm tr \,}

\def\IF{\relax{\rm I\kern-.18em F}}
\def\II{\relax{\rm I\kern-.18em I}}
\def\IP{\relax{\rm I\kern-.18em P}}
\def\IC{\relax\hbox{\kern.25em$\inbar\kern-.3em{\rm C}$}}
\def\IR{\relax{\rm I\kern-.18em R}}

\def\Dsl{\,\raise.15ex\hbox{/}\mkern-13.5mu D} %this one can be subscripted
\def\IZ{Z\kern-.4em  Z}

\catcode`\@=11   
\newdimen\@rotdimen
\newbox\@rotbox  

\def\@vspec#1{\special{ps:#1}}%  passes #1 verbatim to the output
\def\@rotstart#1{\@vspec{gsave currentpoint currentpoint translate
   #1 neg exch neg exch translate}}% #1 can be any origin-fixing transformation
\def\@rotfinish{\@vspec{currentpoint grestore moveto}}% gets back in synch 
\def\@rotr#1{\@rotdimen=\ht#1\advance\@rotdimen by\dp#1%
   \hbox to\@rotdimen{\hskip\ht#1\vbox to\wd#1{\@rotstart{90 rotate}%
   \box#1\vss}\hss}\@rotfinish}
\def\@rotl#1{\@rotdimen=\ht#1\advance\@rotdimen by\dp#1%
   \hbox to\@rotdimen{\vbox to\wd#1{\vskip\wd#1\@rotstart{270 rotate}%
   \box#1\vss}\hss}\@rotfinish}%
\def\@rotu#1{\@rotdimen=\ht#1\advance\@rotdimen by\dp#1%
   \hbox to\wd#1{\hskip\wd#1\vbox to\@rotdimen{\vskip\@rotdimen
   \@rotstart{-1 dup scale}\box#1\vss}\hss}\@rotfinish}%
\def\@rotf#1{\hbox to\wd#1{\hskip\wd#1\@rotstart{-1 1 scale}%
   \box#1\hss}\@rotfinish}%
\def\rotate{\@ifnextchar[{\@rotate}{\@rotate[l]}}
\def\@rotate[#1]#2{\setbox\@rotbox=\hbox{#2}\@nameuse{@rot#1}\@rotbox}

\catcode`\@=12
\begin{document}

\preprint{}

\rightline{ IFT-UAM/CSIC-21-28}

%\begin{frontmatter}
%\title{\Large Constraints on M$_4$ and dS$_4$ vacua\\
%from the  AdS$_3$ non-SUSY swampland conjecture}
\title{\Large AdS Swampland Conjectures\\
and Light Fermions}
\author{Eduardo Gonzalo$^*$}
\author{Luis E. Ib\'a\~nez$^*$}
\author{Irene Valenzuela$^{**}$}
%E. Gonzalo$^*$, L.E. Ib\'a\~nez$^*$  and I. Valenzuela $^{**}$
\address{$^{*}$Departamento de F\'{\i}sica Te\'orica 
and Instituto de F\'{\i}sica Te\'orica  UAM-CSIC,\\
Universidad Aut\'onoma de Madrid,
Cantoblanco, 28049 Madrid, Spain}
\address{$^{**}$Jefferson Physical Laboratory, Harvard University,
Cambridge, MA 02138, USA}
\begin{abstract}
We consider constraints on 
 $D$-dimensional theories in $M_D$, $dS_D$ and $AdS_D$ backgrounds in the light
of AdS swampland conjectures  as applied to their compactification in a circle.
In particular we consider the non-SUSY AdS instability conjecture and the 
AdS distance conjecture. For $M_D$ and $dS_D$ vacua the results may be summarized
by a {\it light fermion conjecture} which states that in  theories with $\Lambda_D\geq 0$ and a
positive  first non-vanishing supertrace $(-1)^{k+1}\text{Str} M^{2k}>0$, a surplus of light fermions with mass
$m_f\lesssim \Lambda_D^{1/D}$  must be present. 
%We argue that the tachyon-free non-SUSY $SO(16)\times SO(16)$ 10D heterotic string is consistent with this conjecture.
%The case of $M_D$ is consistent with the AdS distance conjecture with exponent $\alpha=1/D$. In the $dS_D$ case  consistency of this conjecture requires either light fermions with $m_f\lesssim \Lambda_D^{1/D}$ or that the lightest fermions are part of an emerging  tower of particles.
For $dS_D$ this is supported by both AdS swampland conjectures.
On the contrary, the cases of $M_D$ and $AdS_D$ can be made consistent with the mild but not the strong version of the AdS Distance conjecture, since the KK tower in the lower $d$-dim theory will scale as $m\lesssim \Lambda_d^{\alpha}$ with $\alpha=1/d$.
The above  constraints also suggest that  the Standard Model of particle physics %(and many  $N=1$ sugra extensions ) 
would be inconsistent in Minkowski space but consistent  in dS if the lightest neutrino is Dirac and lighter than the cosmological constant scale.%,as discussed in recent years.

\end{abstract}
\maketitle
%\end{frontmatter}

%\newpage
%%----------------------------------------------------------------------%
%%  Resetting of counters
%%----------------------------------------------------------------------%
%\setcounter{page}{1}
%\pagestyle{plain}
%\renewcommand{\thefootnote}{\arabic{}}
%\setcounter{footnote}{0}
%----------------------------------------------------------------------%
%  Paper begins
%----------------------------------------------------------------------%
%&&&&&&&&&&&&&&&&&&&&&&&&&&&&&&&&&
\section{Introduction}
%&&&&&&&&&&&&&&&&&&&&&&&&&&&&&&&&&

Not every Effective Field Theory (EFT) can be UV completed in a consistent theory of quantum gravity. Recently, a great deal of activity has gone into determining the consistency criteria that allow for this to happen. These conditions are known as swampland constraints and distinguish the landscape of low energy EFTs that can arise from a quantum gravity theory from those that are inconsistent and are said to belong to the Swampland \cite{swampland} (for reviews see \cite{Palti,vafafederico,Irene}).

It is expected that if we start with a  $D$-dimensional theory and compactify it on a circle, if the obtained 
theory in $D-1$ dimensions is inconsistent,  it must be that the original $D$-dimensional theory was
itself inconsistent. Thus a way to test whether a given $D$-dimensional theory is inconsistent is looking 
for an inconsistency in the dimensionally reduced $(D-1)$-dimensional theory. In particular, if the $D$-dimensional theory satisfies some swampland constraint and, it is therefore consistent with quantum gravity, then the lower dimensional theory should do as well. This line of reasoning has been very useful in the past  to promote some string theory feature to a general quantum gravity principle, rather than simply being an artefact of the lamppost we look under.  In particular, it has been used to argue for the ubiquitous presence of Chern-Simons terms  in order to avoid the presence of global symmetries in lower dimensions \cite{Montero:2017yja}, and certain properties of the gauge kinetic function in order to avoid the presence of 2d deSitter vacua \cite{Montero:2020rpl}. It has also been used to refine certain swampland conjectures to make them robust under dimensionally reduction, like the Weak Gravity Conjecture \cite{WGC} in \cite{Heidenreich:2015nta} as well other more recent swampland conjectures in \cite{tom}.

In this paper, we will impose the following swampland constraints:

\begin{itemize}
\item Non-SUSY AdS conjecture \cite{OV,Freivogel:2016qwc}: Any non-supersymmetric vacuum must be at best metastable.
\item AdS Distance conjecture \cite{Lust:2019zwm}: There must exist an infinite tower of states with mass 
\beq 
\label{ADC}m^{(D)}_{n} \sim n|\Lambda_{D}|^{\alpha_{D}}M_{D}^{1-2\alpha_{D}}\eeq
 as $\Lambda_D\rightarrow 0$ when scanning a family of vacua with cosmological constant $\Lambda_{D}$ .  The mild version implies $\alpha_D$ to be a positive constant, while the strong version requires $\alpha_D\geq 1/2$ for AdS vacua.

\end{itemize}

Consider some Einstein gravity theory in $D$ dimensions coupled to matter satisfying the above swampland constraints. Our goal is to determine the constraints on the spectra such that circle compactifications of this theory preserve the above swampland criteria as well. To answer this question, we will study the potential for the radion arising in lower dimensions from the Casimir Energy contribution of the different states. If we find that
the dimensionally reduced theory leads to a vacuum violating the above swampland constraints, the $D$-dimensional theory
should be inconsistent.
This will allow us to rule out certain universal classes of Minkowski, dS and AdS vacua. %In the case of Minkowski and dS we find that the presence of light fermions of mass $m_f\lesssim \Lambda^{1/d}$ in the spectrum is enough to prevent this theories from falling in the Swampland. 

Our results hold under the following assumptions.
%\begin{itemize}
First, when applying the Non-SUSY AdS conjecture, we will assume that potential non-perturbative instabilities in $D$ dimensions are not inherited by the AdS $(D-1)$-dimensional vacua. This implies that the bubble radius of these instabilities (if it exists) is larger than the AdS length, so they cannot describe bubble instabilities in lower dimensions. This is better justified the smaller the AdS length is.

Secondly, when applying the AdS Distance conjecture, we will assume that there is a family of $D$-dimensional vacua exhibiting different values for the masses of the states in such a way that by scanning on the masses we are effectively probing different EFTs. We will assume that the point where all states are massless is part of the landscape and start increasing the masses until we reach (if so) some inconsistency. 
%\end{itemize}

A common feature found in our analysis is the need for light fermions in the theory. In particular that is the case for Minkowski and dS vacua and 
is supported  by  the results of both the Non-SUSY AdS conjecture and the AdS distance conjecture. 
The fact that two independent swampland constraints require the presence of light fermions is intriguing and suggests that
they are signals of a more general quantum gravity constraint involving light fermions.
In particular we propose a 
{\it light fermion conjecture } which states that in theories with $\Lambda_D\geq 0$ and positive  first non-vanishing supertrace $(-1)^{k+1}{\cal S}tr M^{2k}>0$,
there must be a surplus of fermions with masses $m\lesssim \Lambda_D$.  
%We argue that the non-SUSY,  $SO(16)\times SO(16)$ heterotic string is
%an example consistent with this conjecture. 
The Standard Model (SM) itself may be an example if one interprets dark energy in terms of a dS phase,
with the predicted light fermions being the lightest neutrino generation. In fact, this leads to the prediction that the lightest neutrino must be Dirac and have a
mass $m_{\nu_1}\lesssim \Lambda_4^{1/4}$, which nicely connects two fundamental scales of physics and is consistent with present neutrino data.
This has been studied in detail in \cite{OV,IMV1,HS,2toro,Gonzalo} and in a companion paper which also studies the case in which the dS phase is due to a quintessence scalar \cite{companion}.

\section{Setup}

Consider a ($D>3$)-dimensional Einstein gravity theory coupled to matter. Throughout this note, we will assume the following $D$-dimensional spectra:
\begin{itemize}
\item Massless graviton + massless $U(1)^a$ gauge bosons. This yields $n_0=2(a+1)$ massless bosonic degrees of freedom.
\item Massive scalar bosonic degrees of freedom.
\item Massive fermionic degrees of freedom.
\end{itemize}
We will always denote as $n_b$ ($n_f$) the total number (massless + massive) of bosonic (fermionic) degrees of freedom.
We can trivially extend the results to more massless fields by simply interpreting $n_0$ as the net number of massless bosonic degrees of freedom (i.e. massless bosons minus massless fermions). 

Upon dimensional reduction on a circle, we obtain a $d=D-1$ dimensional theory including at least one extra scalar, the radion $R$, which parametrizes the size of the circle,
\begin{equation}
2\pi R=\int_{0}^{2\pi r}\sqrt{G_{DD}}dx^{D}\label{radiondefine}\ .
\end{equation}
 To set notation, we choose the following parametrization for the metric
\begin{equation}
G_{MN}=\left[\begin{array}{cc}
(R/r)^{\frac{-2}{(d-2)}}g_{\mu \nu } & 0\\
0 & (R/r)^{2}
\end{array}\right],\label{metric}
\end{equation}
where $r$ is introduced to make the metric dimensionless.

The potential for the radion includes a tree-level contribution from a possible $D$-dimensional cosmological constant $\Lambda_D$ as well as a one-loop piece originated from the Casimir energy contribution of all the states of the higher dimensional theory. This reads:
\beq
V=V_{\rm tree}+V_{1L}
\eeq
where 
\begin{equation}
V_{\text{tree}}=2\pi r\left(\frac{r}{R}\right)^{\frac{2}{d-2}}\Lambda_{D}
\label{eq:tree}
\end{equation}

\begin{align}
\label{casimir}
V_{\text{1L}} & =\pm2\pi r^{\frac{(D-1)}{(D-3)}}\sum_{i=b,f}\frac{n_{b,f}\,\,2m_{i}^{D}}{(2\pi)^{\frac{D}{2}}R^{\frac{2}{(D-3)}}}\nonumber \\
& \times\sum_{n=1}^{\infty}\frac{K_{D/2}(2\pi nm_{_{i}}R)}{(2\pi nm_{i}R)^{D/2}},
\end{align}
where $K$ is a modified Bessel function of the second kind.
The plus sign is for fermions (with periodic boundary conditions) while the minus sign is for bosons. This one-loop effective potential can be computed
by a Gaussian integral in the path integral formalism using the
background field method. The computation of the Casimir energy in $M_{d}\times S_{1}$ can
also be found in \cite{ArkaniHamed:2007gg}. 

In the region $R\ll(2\pi nm_{_{i}})^{-1}$ the potential of a single
degree of freedom is given by:

\begin{equation}
V_{1L}\approx\frac{(-1)^{F}r^{\frac{(D-1)}{(D-3)}}}{R^{\frac{(D-1)(D-2)}{(D-3)}}}\sum_{k<\frac{D}{2}}\left(-1\right)^{k+1}\beta_{k}(mR)^{2k}
\label{expansion0}
\end{equation}
where $\beta_{k}=\frac{\Gamma(\frac{D}{2}-k)\zeta(D-2k)}{2^{D-1}\pi^{D-1}\Gamma(k+1)\pi^{\frac{D}{2}-2k}}$. This expansion is obtained by introducing an asymptotic formula for the modified Bessel function before making the infinite summation. It is only valid up to $2k<D$, when different terms would have to be included.

Summing over all degrees of freedom, we get that the potential as $R\rightarrow 0$ behaves to leading order as
\begin{equation}
V(R\rightarrow 0)\approx\frac{r^{\frac{(D-1)}{(D-3)}}}{R^{\frac{(D-1)(D-2)}{(D-3)}}}\sum_{k<\frac{D}{2}}\beta_{k}(-1)^{k+1} \text{Str}(M^{2k})R^{2k}+\dots
\label{expansion}
\end{equation}
where we have defined the supertrace of $M^{2k}$ as
\beq
 \text{Str}(M^{2k})=\sum_{b}n_{b}m_{b}^{2k}-\sum_{f}n_{f}m_{f}^{2k}\ .
 \label{supertrace_defined}
\eeq
Hence, the potential in the UV (as $R\rightarrow 0$) goes to $\pm \infty$ where the sign is determined by the first non-zero supertrace $V\propto (-1)^{k+1} \text{Str}(M^{2k})$ with $k=0,1,2,\dots$
Notice that the first supertrace with $k=0$ is simply equal to the difference between the total number of fermionic and bosonic degrees of freedom, so that the sign is determined by  $(-1)\text{Str}(M^0)=\sum_{f}n_{f}-\sum_{b}n_{b}$. If this term cancels out due to an equal number of bosons and fermions, then the sign of the potential
is rather determined by 
$
 \text{Str}(M^2)=\sum_{b}n_{b}m_{b}^{2}-\sum_{f}n_{f}m_{f}^{2}
$, and so on.
%If the supertrace of $M^2$ is also zero, then it is determined by $(\sum_{f}n_{f}m_{f}^{4}-\sum_{b}n_{b}m_{b}^{4})$ etc.. In general, in the UV goes to $\pm \infty$ and this sign is determined by the first non-zero supertrace $V\propto (-1)^{k+1} \text{Str}(M^{2k})$ for $k=0,1,2$… 

In the region $R\gg(2\pi nm_{_{i}})^{-1}$ the modified Bessel functions
of the second kind decay exponentially, so only the massless particles
in the $D$ dimensional spectra contribute. Therefore, unless
there is a surplus of massless fermions in the spectra, the graviton (and other
massless bosonic mediators) dominate the one-loop potential in the
deep IR. This means that the Casimir potential goes to zero from negative
values when $R\rightarrow\infty$. On top of this, one must consider the tree level contribution, which for $\Lambda_D \neq 0$ dominates over the Casimir contribution and controls the behaviour of the potential  for large radius, so that the complete potential goes to zero from (negative) positive values for (A)dS. More concretely, we can write the leading order expansion of the potential for $R\rightarrow \infty$ as
\begin{equation}
V(R\rightarrow \infty)\approx2\pi r\left(\frac{r}{R}\right)^{\frac{2}{d-2}}\Lambda_{D} -n_0\frac{(-1)^{F}r^{\frac{(D-1)}{(D-3)}}\beta_{0}}{R^{\frac{(D-1)(D-2)}{(D-3)}}} 
\label{expansion3}
\end{equation}
where recall that $n_0$ is the net number of massless bosonic degrees of freedom.

In the following, we will study in detail under which conditions of the $D$-dim spectra, the above radion potential generates a $d$-dim AdS vacuum  that may violate the AdS swampland conjectures. Notice that, in the presence of massless $U(1)$'s 
 additional massless scalar Wilson-line degrees of freedom may appear in the spectrum. Their presence may give rise to non-perturbative instabilities of the $d$-dimensional radion AdS vacua we are studying \cite{HS}. When discussing the Non-SUSY AdS conjecture we will assume that such 
non-perturbative instabilities are not present. This may be due to the nucleation dynamics. In particular the AdS vacua will be stable if the
bubble radii are larger than the AdS length so that the decay does not proceed \cite{HS}. The possibility of the existence of any vacua of these characteristics will
make the AdS stability conjecture to apply. 
Alternatively one can consider 
closely related compactifications on the segment ${\cal S}^1/Z_2$ with the $Z_2$ projecting out the Wilson line scalars \cite{2toro}. If that is done,  all the computations  and the discussion below remain essentially  identical except for overall 1/2 factors. 
For simplicity we will only discuss the case of circle compactification and ignore this potential  instability in what follows.
It is also important to remark  that such potential instabilities  would not affect the constraints coming from the AdS distance conjecture, which do  not depend on the
obtained  AdS minima in the compactified theory being local or global. 

\begin{figure}
	\centering{}
	\includegraphics[scale=0.25]{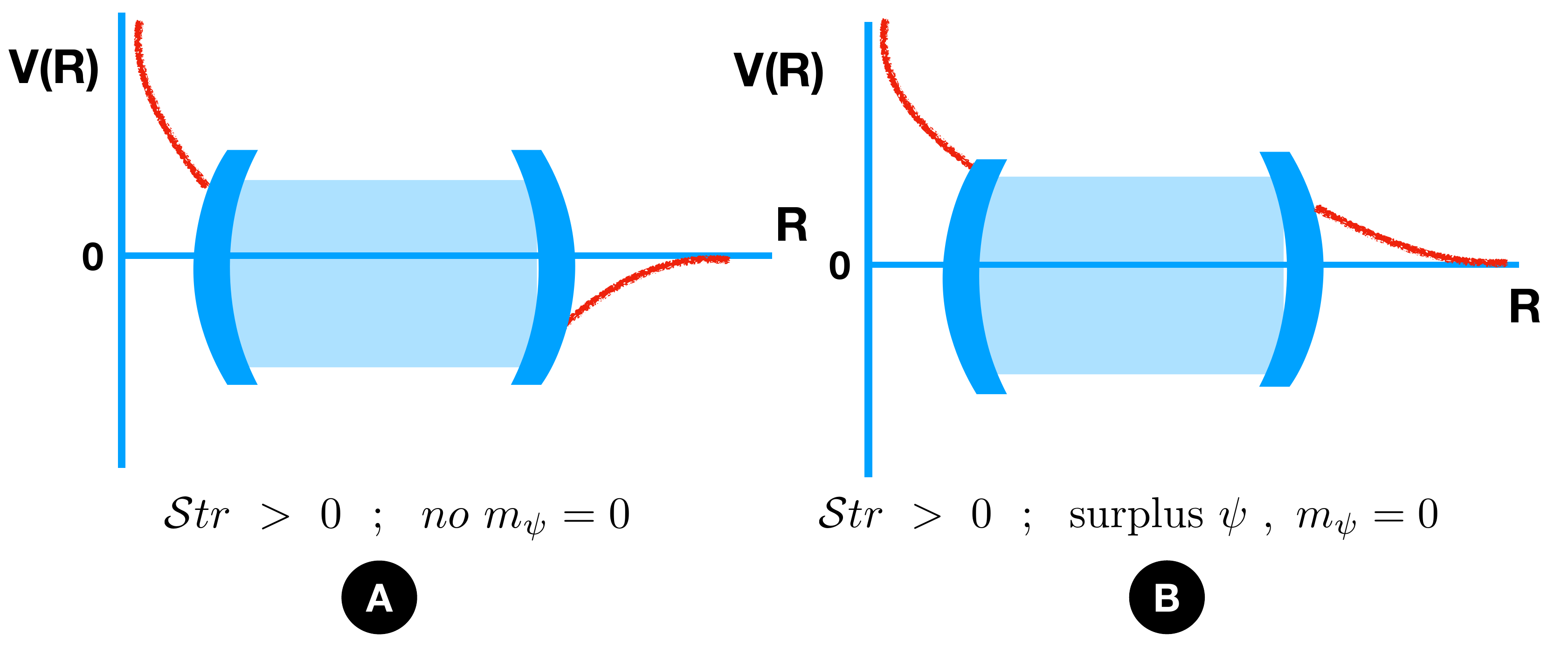} 
	\caption{\footnotesize Schematic representation of the radion $(D-1)$-dimensional potential from  $D$-dimensional Minkowski vacua. Case A is in the swampland on the
	grounds of the non-SUSY AdS conjecture. Here  ${\cal S}tr>0$ stands for a theory satisfying $(-1)^{k+1} \text{Str} \, (M^{2k})>0$, $k=0,1,2...$ for the first non-vanishing supertrace. In Case B a surplus of massless fermions may render the theory consistent.}
	\label{fig:Mink1}
\end{figure}

\section{Constraints on Minkowski vacua}

\subsection{From Non-SUSY AdS Conjecture}

The form of the radion potential in $d$ dimensions is sensitive to  the number of bosons and fermions, $n_b,n_f$ as well
as their masses. Let us consider first the case in which the only massless particles are the graviton and possibly 
massless $U(1)^a$ gauge bosons. In this case, for large $R$, the Casimir potential goes to zero 
from negative values. On the other hand, if there are more fermions than bosons $n_f>n_b$, for small R the
potential is dominated by the first term  ($k=0$) in  \eqref{expansion} which will diverge like $R^{-\frac{(D-1)(D-2)}{(D-3)}}$, so the potential becomes positive at small radius.  Thus, we conclude that  an AdS vacuum will develop somewhere  in between, see Fig.\ref{fig:Mink1}.   Since it has more fermions than bosons, the $D$-dim theory 
was not SUSY, so neither is the $d$-dim theory. Therefore, if these lower dimensional AdS vacua are stable, they would violate the Non-SUSY AdS swampland conjecture. Notice that we are choosing periodic fermionic boundary conditions, so that the Witten {\it bubble of nothing} \cite{witten} is not topologically allowed. There could be, though, some UV spin defect that allow a bubble of nothing to form, as required by the swampland conjecture in \cite{McNamara:2019rup} regarding triviality of the cobordism group. However, in that case, the bubble instability is expected to be highly suppressed, implying that the bubble radius is very large. If this bubble radius is larger than the AdS length, it will not describe a bubble instability in lower dimensions and the AdS vacua will be stable. We will take this as an assumption, which bring us to the conclusion that the the original D-dimensional Minkowski vacua with $n_f>n_b$ are inconsistent with quantum gravity. The same result applies if we have massless fermions, unless the number of massless fermionic degrees of freedom exceeds the bosonic ones, which would change the sign of the potential at large radius (see Fig.\ref{fig:Mink1}).

Let us consider now the case of $D$ Minkowski vacua with equal number of fermions and bosons, $n_f=n_b$.
This is necessary, for instance, if the Minkowski vacua restore supersymmetry at some energy scale. 
Due to the presence of the massless graviton plus possibly some gauge fields, the potential still goes to zero from negative values  for large R. As $R\rightarrow 0$, the leading term of the expansion of the potential in \eqref{expansion} identically vanishes since $n_f=n_b$, so the potential is dominated by the  term  proportional to the supertrace of $M^2$.
If the latter is positive,  a global AdS minimum must develop at some intermediate value of R, which would be inconsistent with the Non-SUSY AdS conjecture. If not only $n_f=n_b$ but also the supertrace of $M^2$ is zero, then the potential is dominated by the first non-vanishing supertrace $Str(M^{2k})$. Notice that the sign alternates between the different terms, so for example an AdS runaway will appear if $Str(M^{4})$ is positive.% This change of sign follows from Eq. (\ref{expansion}). 

As we said,
one way out to avoid the $d$-dim AdS vacuum is to add  massless fermions (at least as many as massless bosons) to change the trend of the potential at large radius. If the fermions dominate,
the potential will be positive at large $R$, and no AdS vacua will be necessarily generated.
Massless fermions may appear in different instances. In particular they may appear if there is an anomaly-free set of  chiral fermions
charged under massless (Abelian or non-Abelian) vector bosons.
One may also consider massless fermions with masses protected by e.g. discrete gauge symmetries.
This brings us to the following statement:

\begin{center}
{\bf Claim 1:} {\it A $D$-dim Minkowski theory satisfying $(-1)^{k+1} \text{Str} \, (M^{2k})>0$ for the first non-vanishing supertrace is
inconsistent with quantum gravity unless there is a surplus of massless fermions.}
\end{center}

In particular, this rules out Minkowski vacua in non-supersymmetric theories with $n_f>n_b$ and theories with supersymmetry broken spontaneously (so $n_f=n_b$) with $\text{Str} \, (M^{2})>0$, as long as the massless degrees of freedom are mainly bosonic.

Interestingly, there are well known explicit formulae for the supertrace in ${\cal N}=1$ $D=4$ supergravity 
theories with SUSY spontaneously broken at tree level.  The supertrace
at the minimum of a general supergravity potential is given by (see e.g.  \cite{Ferrara} and references therein)
\beq
\begin{split}
 \text{Str}\  ({\cal M}^2)\ =\ 2(N-1)\ \left(V_0\ +\ |m_{3/2}|^2\right)\ +\\
+\  2e^{\cal K}\ 
{\cal R}_{\alpha {\bar \beta}} ({\overline D}^\alpha {\overline W} D^{\bar \beta}W) \ .
\end{split}
\label{supertrace}
\eeq
Here {\cal K} is the Kahler potential, $W$ the superpotential an ${\cal R}_{\alpha {\bar \beta}}$ the 
curvature tensor associated to the scalar metric. Also, $N$ is the number of chiral multiplets
and $V_0$ the value of the potential at the minimum, which vanishes in Minkowski.
There are some prominent examples in which the supertrace is positive, which could therefore risk to be inconsistent with quantum gravity in $M_4$. 
In particular, if the scalar metric is canonical (sometimes called minimal ${\cal N}=1$ sugra
in phenomenological applications) the curvature piece vanishes and $ \text{Str} ({\cal M}^2)>0$ for more than one chiral multiplet $N>1$.
Such class of theories would then be in the swampland unless some massless fermions exist.
On the contrary, in no-scale supergravity theories with $N$ chiral multiplets one has ${\cal R}_{\alpha {\bar \beta }}\ =\ -\ g_{\alpha {\bar \beta}}(N+1)/3$ where $g_{\alpha\bar\beta}$ is the field metric.
By plugging this back into Eq. (\ref{supertrace}) and using the $\mathcal{N}=1$ supergravity formula for the potential we get that $\text{Str} ({\cal M}^2)=-4|m_{3/2}|^2<0$. Hence, no AdS
vacuum develops in general and such models cannot be excluded on these grounds.

Clearly, the case with $n_f=n_b$ is not restricted to supersymmetric setups but can also occur in the absence of supersymmetry. However, if the theory is non-SUSY we cannot guarantee that the first more divergent 
${\cal O}(R^{-\frac{(D-1)(D-2)}{(D-3)}})$  term
vanishes to all orders  in perturbation theory in the computation of the Casimir potential, as happens in the SUSY case. 
We would need to go beyond the 1-loop vacuum energy in \eqref{casimir} to determine the behaviour of the potential as $R\rightarrow 0$.
Hence, we cannot
ascertain without further calculation whether an AdS vacuum develops or not for a non-SUSY theory with $n_f=n_b$.

Finally, note that the Claim 1 above implies that the SM as such would be in the swampland,  if we were in
Minkowski space.  Indeed the number of Weyl fermions in the SM is 48 (including right-handed neutrinos) whereas there are 12 gauge bosons
and  4 Higgs scalars, so $n_f>n_b$. Furthermore, we only have massless bosonic degrees of freedom. We will see momentarily that this conclusion may be evaded if there is 
a non-vanishing positive vacuum energy. So using these ideas one could have predicted before the actual discovery that
a non-vanishing cosmological constant (or some form of quintessence) must exist in the present universe. This could be avoided if there is one generation of strictly massless Dirac neutrinos, e.g. charged under some discrete gauge symmetry protecting them from becoming massive.
In this case the 4 neutrino degrees of freedom would exactly cancel the graviton+photon contributions. %We refer the interested reader to the refs \cite{OV, IMV1,HS,2toro} and a companion  paper to appear \cite{companion}.

\subsection{From AdS Distance Conjecture}
Unlike what happens with the previous conjecture, here non-susy AdS vacua are not problematic per se, but only if they are part of a family of vacua of different vacuum energy such that $\Lambda_d$ can be taken to be parametrically small without an infinite tower of massless becoming light. As we have seen, upon compactification on a circle,
the $d-$dimensional cosmological constant $\Lambda_d$ depends on the masses of the $D$-dimensional fields. Hence, in order to determine whether the original vacuum is inconsistent with the AdS Distance conjecture,  we need to check whether by varying these masses we can send $\Lambda_d \rightarrow0$ without forcing an infinite tower of states to become light in the limit. We will therefore assume that such scanning of the masses is possible in the sense that the original vacuum is not isolated but rather forms part of a family of $D$-dim Minkowski vacua exhibiting different values of the masses. The advantage is that the results are independent of whether the vacuum is unstable or supersymmetric, and therefore no further assumption about the UV instability of the $D$-dim vacuum is required. In particular it is independent of the existence of other minima or runaway directions in the UV.

\begin{figure}
	\centering{}
	\includegraphics[scale=0.25]{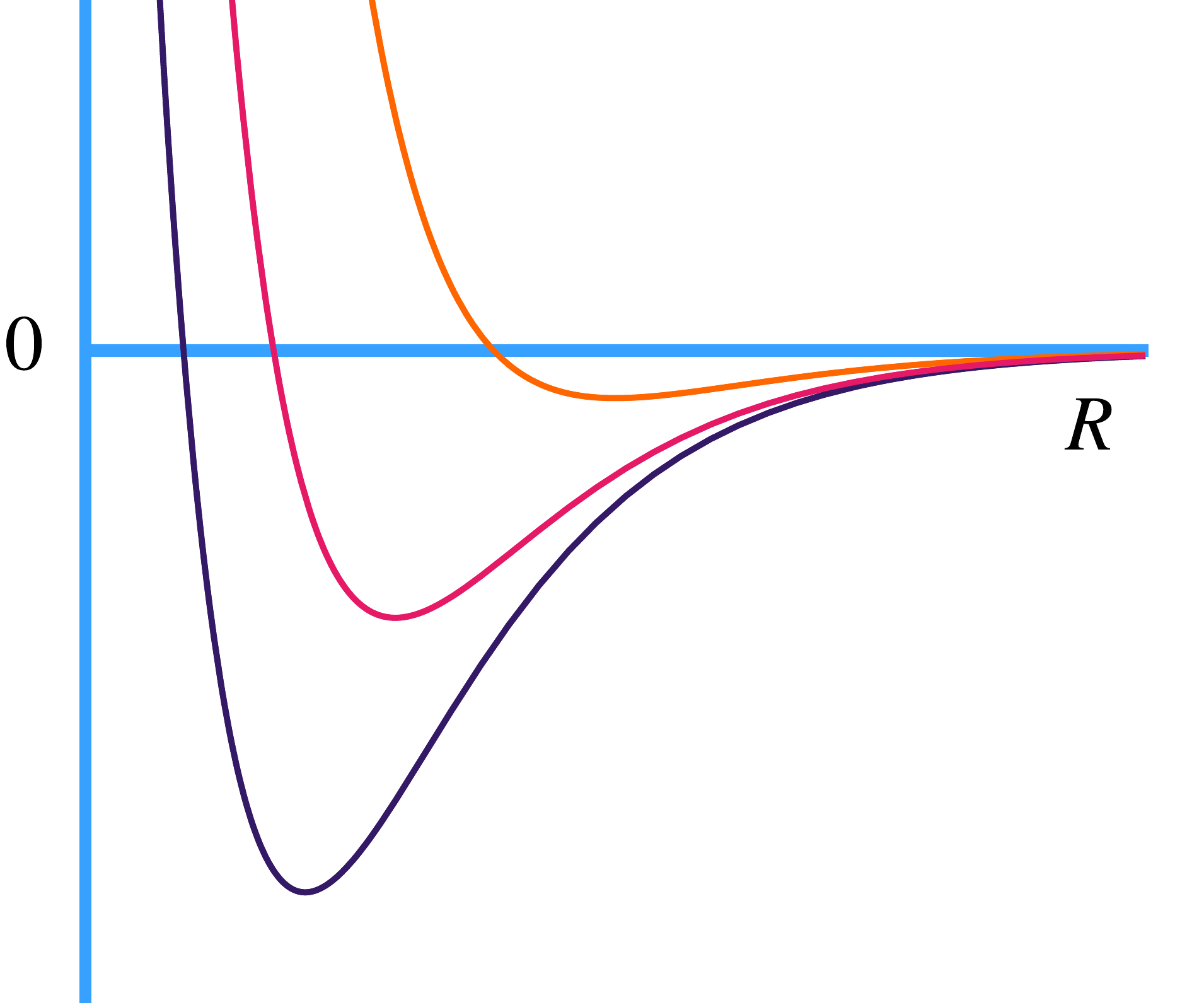} 
	\caption{\footnotesize Schematic representation of the radion $(D-1)$-dimensional potential from  $D$-dimensional  Minkowski vacua. We find that, as the fermions become lighter, $R_{0}\rightarrow\infty$, $\Lambda_{d}\rightarrow0$ and that the KK tower becomes light as $M_{KK}\propto \Lambda_{d}^{\frac{1}{d}}$. Therefore, we find agreement with the mild AdS Distance conjecture. }
	\label{fig:Mink2}
\end{figure}

Let us assume that an AdS vacuum is generated in $d$-dimensions. One can check that minimization of the potential implies:

\begin{align}
\sum_{i=b,f}\frac{(-1)^{F}n_{i}m_{i}^{D+\frac{2}{D-3}}}{(m_{i}R_{0})^{\frac{2}{D-3}}}\biggl\{\frac{2}{D-3}\times\sum_{p=1}^{\infty}\frac{K_{\frac{D}{2}}(2\pi pm_{i}R_{0})}{(2\pi pm_{i}R_{0})^{\frac{D}{2}}}\nonumber \\
+\sum_{p=1}^{\infty}\frac{pK_{\frac{D}{2}+1}(2\pi pm_{i}R_{0})}{(2\pi pm_{i}R_{0})^{\frac{D}{2}-1}}\biggr\} & =0,
\end{align}
where $R_0$ is the value of the radius at the minimum.
If all the massive particles have the same mass $m$ and if such a minimum exists, then it should scale as $R_{0}\propto\frac{1}{m}$, since the above expression is a function $f(mR_0)$ of the product of the mass and the radius. By plugging this back into \eqref{casimir}, we can see that the potential at the minimum will scale as 
\beq
\Lambda_d\simeq V_{0}\propto r^{\frac{D-1}{D-3}}R_{0}^{-(D+\frac{2}{D-3})}=(r^{\frac{1}{d-2}}R_{0}^{-(\frac{d-1}{d-2})})^{d}
\eeq
where $D=d+1$ and we have set $M_P=1$.  Hence, in the limit $m\rightarrow 0$, $R_{0}\rightarrow\infty$ and $V_0\rightarrow 0$ (see Fig.\ref{fig:Mink2}). 
%\vspace{-0.06cm}Defining $x=mR_{0}$ this can be rewritten as $f(x)=0.$ 
%In the cases
%where there is a minima, this equation will have a solution $x=x_{0}$.
%Therefore, at the minima $R_{0}\propto\frac{1}{m}.$ Thus, as we scan
%along $m\rightarrow0$ ($R_{0}\rightarrow\infty$), the minima changes
%as $V_{0}\propto r^{\frac{D-1}{D-3}}R_{0}^{-(D+\frac{2}{D-3})}=(r^{\frac{1}{d-2}}R_{0}^{-(\frac{d-1}{d-2})})^{d}$.

According to the AdS Distance conjecture, an infinite tower should become light as $V_0\rightarrow 0$. Indeed, the Kaluza Klein mass is given by $M_{KK}^{2}=\sqrt{G}\frac{n^{2}}{r^{2}}G^{DD}$,
where $G_{DD}$ is the the $D$-th component of the space-time metric. %$\frac{n^{2}}{r^{2}}G^{DD}$ come from acting with $G^{DD}\partial_{D}\partial_{D}$on $e^{\frac{inx^{D}}{r}}$.
 In our chosen parametrization in Eq. \eqref{metric}, one gets $M_{KK}=n\,r^{\frac{1}{d-2}}R^{-(\frac{d-1}{d-2})}$
%(it has the right units since both $r$ and $R$ has units of length)
which implies that
\beq 
M_{KK}\sim nV_{0}^{\frac{1}{d}} \ .
\eeq
Therefore, there is indeed a tower (KK modes) becoming light as $V_0\rightarrow 0$ and scaling as required by the AdS Distance Conjecture, namely $m_{\rm tower}\sim \Lambda_d^\alpha$ with $\alpha=\frac{1}{d}$.
Since the limit $\Lambda_d\rightarrow0$ implies the limit
$R\rightarrow\infty$, the AdS distance conjecture coincides with
the standard Distance conjecture in the moduli space, and it is therefore ``trivially'' satisfied thanks to the KK tower.

We could also consider the more general case where we have several
particles in $D$ dimensions whose masses $m_{i}$ are not necessarily the same.
In order to apply the AdS distance conjecture we first need to specify a particular scanning trajectory in the masses that allows us to take the limit $\Lambda_d\rightarrow 0$. Of particular interest is the case in which
all of the masses change at the same rate. This is what happens for
example in the Standard Model, where the masses of the particles all
depend on the vev of the same field, the Higgs. It is also natural if all masses are generated upon breaking supersymmetry and we vary the supersymmetry breaking scale. Then, in the limit
$m_{i}\rightarrow0$ the position of the minima
goes to $R\rightarrow\infty$ while $\Lambda\rightarrow0$ and the
KK towers come down again as $M_{KK}\propto n\Lambda^{\frac{1}{d}}$.

To conclude, whereas the Non-SUSY AdS conjecture is very predictive constraining Minkowski vacua, the AdS distance conjecture is automatically consistent for Minkowski vacua as one cannot take the limit $\Lambda_d\rightarrow 0$ without increasing the radius and making the KK tower light. It is worth noting, though, that the conjecture is satisfied with an exponent $\alpha=\frac{1}{d}$. This is in disagreement with the strong version of the conjecture requiring $\alpha\geq\frac{1}{2}$ with saturation for the supersymmetric case \cite{Lust:2019zwm}. The same exponent $\alpha=1/d$ was recently proposed in \cite{tom} by different means. Hence, either only the mild version of the conjecture is correct, or these Minkowski vacua are inconsistent with quantum gravity. If we take the latter perspective and impose the strong version of the conjecture, then the Minkowski vacua that are inconsistent are those generating any local AdS minimum in the compactified theory, which includes those that are also inconsistent with the Non-SUSY AdS conjecture highlighted in Claim 1. Again, only a surplus of massless fermions would avoid the formation of AdS vacua. %Hence in this case the AdS distance conjecture does not lead to additional constraints.

\section{Constraints on de Sitter vacua}
\subsection{From Non-SUSY AdS Conjecture}

Let us consider now the case in which the starting theory in $D$-dimensions is in dS space. It has been 
conjectured that dS vacua are in the swampland \cite{Obied:2018sgi,Dvali:2013eja,Dvali:2014gua,Dvali:2017eba}, or at least that their lifetime must be smaller than a Hubble time  \cite{Bedroya:2019snp}.  In what follows we will assume that metastable dS vacua exist, but that the possible AdS $d-$dimensional vacua obtained upon compactification are nevertheless stable. As discussed previously, this is possible if the bubble instability mediating the decay in dS is too large to fit in the lower dimensional AdS vacuum. Although admittedly this is not a mild assumption, it is interesting to discuss the consequences that emerge from this assumption as it can explain some of the naturalness issues observed in our universe, as pointed out in \cite{IMV2,Gonzalo}. Moreover, we expect some of our results (in particular, the requirement of light enough fermions) to remain true even in a quintessence scenario replacing the dS vacuum, as we show for the case of the Standard Model in \cite{companion}. Interestingly, the same results will also be obtained when studying the constraints from the AdS Distance conjecture in the next subsection.

\begin{figure}
	\centering{}
	\includegraphics[scale=0.25]{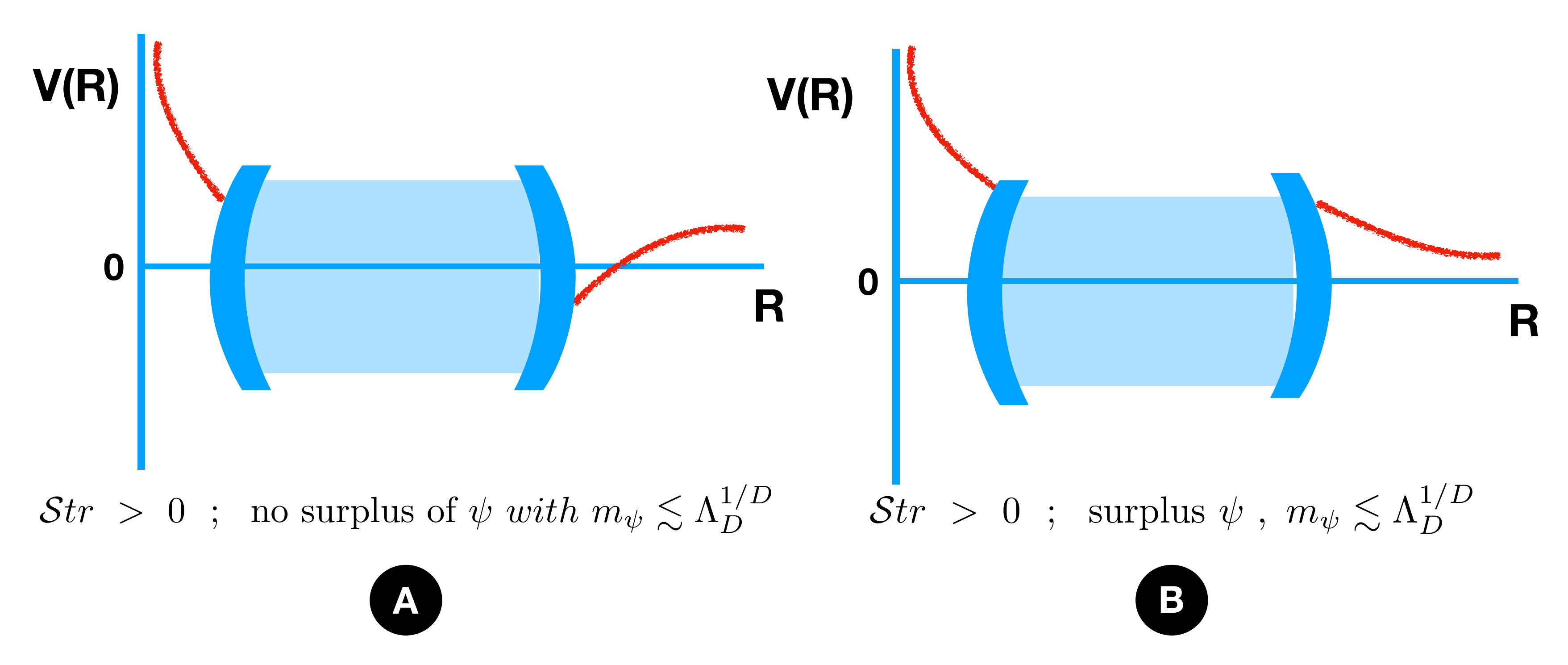} 
	\caption{\footnotesize Schematic representation of the radion $(D-1)$-dimensional potential from  $D$-dimensional  dS vacua. The case  A is  in the swampland,
		on the
		grounds of the non-SUSY AdS conjecture. In the presence of a surplus of fermions with $m_\psi\lesssim \Lambda^{1/D}$ it may become consistent (case B).}
		\label{fig:dS1}
\end{figure}

In the case of a pre-existent constant positive vacuum energy in $D$ dimensions the situation changes significantly compared to the Minkowski case above. The crucial difference is that, at large $R$, it is the tree level dimensional reduction
of the $D$-dimensional c.c. Eq. \eqref{eq:tree} which dominates and provides a {\it positive} contribution to the potential behaving like ${\cal O}(R^{-\frac{2}{D-3}})$ at large radius.
Then if there are more massless bosonic than fermionic degrees of freedom, as naturally expected due to the existence of the graviton (with possibly additional massless $U(1)$'s),  the potential will eventually decrease as 
$R$ decreases and become negative at some point  $R\sim (\Lambda_D)^{-1/D}$. Then,  in total analogy with the
Minkowski case, AdS vacua will form if the first non-vanishing supertrace is positive, $(-1)^{k+1} \text{Str} \, (M^{2k})>0$. This includes the non-supersymmetric case with $n_f>n_b$ or the case of supersymmetry spontaneously broken with $n_f=n_b$ and positive supertrace (see Fig.\ref{fig:dS1}). 
However this may be avoided if there are at least $(1+n)$ Weyl  sufficiently light fermions. Unlike in Minkowski, these fermions do not need to be strictly massless, but one can check that it is sufficient to have
$m_i\lesssim \Lambda_D^{1/D}$ to avoid the potential to become negative. In such a case, the potential might still develop some dS minimum, but it will always be metastable due to the runaway behaviour to large radius and, therefore, consistent with the Non-SUSY conjecture. 
Thus one can make the following general statement:

\begin{center}
	{\bf Claim 2:}   {\it A $D$-dimensional  de Sitter vacua satisfying $(-1)^{k+1} \text{Str} \, (M^{2k})>0$ for the first non-vanishing supertrace is inconsistent with quantum gravity unless there are enough light fermions with mass $m\lesssim \Lambda_{D}^{1/D}$.}
	\end{center}

%Notice that the case $k=0$ is simply saying $n_f>n_b$. 
Let us remark that enough light fermions means that the number of fermionic minus bosonic degrees of freedom with masses below the cosmological constant has to be positive. 

The reader may note that the SM  of particle physics along with the observed non-vanishing c.c. may fit with the Claim 2 above
if we identify the required light Weyl fermions with some neutrino degree of freedom. 
 In fact, one needs 2 Weyl fermions, due to the existence of the massless photon to cancel the negative Casimir contribution from the graviton and the photon and avoid the AdS minimum. Thus, as we said, a Dirac neutrino with mass
 $m_{\nu_1}\lesssim \Lambda_{D}^{1/D}$  would be sufficient to guarantee consistency with the swampland conjecture. 
 This has been discussed at length in \cite{OV,IMV1,HS,2toro}. Amazingly, it could explain the numerical coincidence between neutrino masses and the cosmological constant observed in our universe and provide a new insight into the EW hierarchy problem as first pointed out in \cite{IMV2,Gonzalo}.

 It is also interesting to note what happens in the case of a ${\cal N}=1$ version of the SM. In this case, an AdS vacuum can be avoided either if the lightest neutrino is sufficiently light or
 if the supertrace is non-positive $\text{Str} \,{\cal M}^2\leq 0$ implying a runaway behaviour towards $V_d\rightarrow -\infty$ for small radius. This latter condition on the supertrace is a strong constraint
 on SUSY versions of the SM.  For example, in Split SUSY \cite{Nima,Gian}  the gauginos and Higgsinos are much ligther than
 the sfermions. Thus, the supertrace will be positive and a light neutrino is needed to avoid the presence of AdS$_d$ vacua.
 Of course, this assumes that the SUSY-breaking and invisible sectors of the theory do not contribute substantially to
 the supertrace.

 \subsection{From AdS Distance Conjecture}

\begin{figure}
	\centering{}
	\includegraphics[scale=0.25]{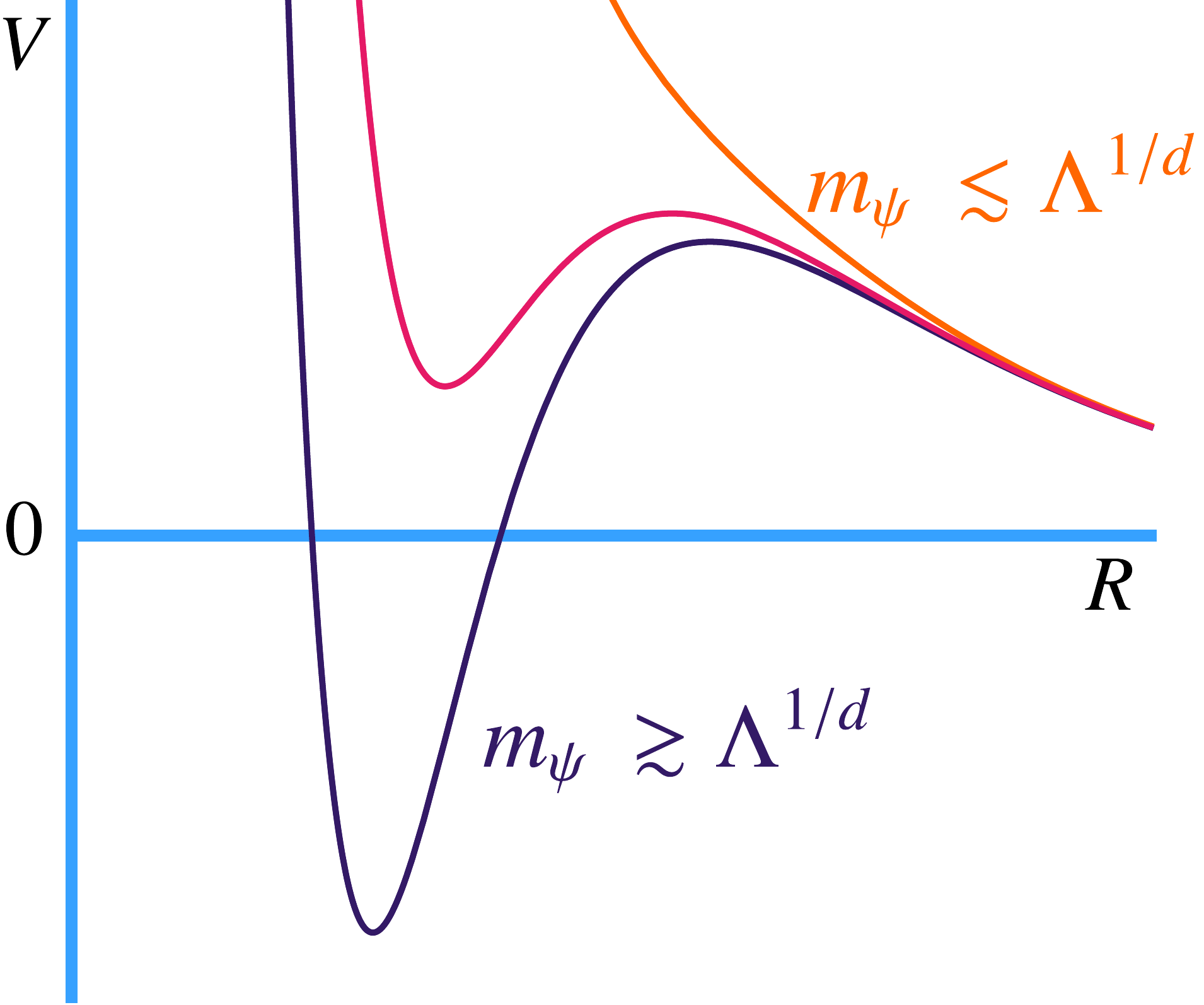} 
	\caption{\footnotesize Schematic representation of the radion $(D-1)$-dimensional potential from  $D$-dimensional  dS vacua. We find that, as the fermions become heavier,  $\Lambda_{d}\rightarrow0$ (and then it becomes negative) at a finite value $R_{0}$. We don't expect a tower to become light at that particular value of $R_{0}$ so we find a disagreement with the mild AdS Distance conjecture unless the fermionic mass is bounded by the scale set by the $D$-dimensional cosmological constant.}
		\label{fig:dS2}
\end{figure}

 As explained above, the $D$-dimensional scalar potential goes to zero as $R^{-\frac{2}{D-3}}$ for large radius due to the positive $D$-dim cosmological constant. As R decreases, it generates a maximum and starts decreasing due to the contribution from the massless bosonic fields (graviton plus possibly gauge fields). Hence, a minimum will be generated if we have $n_f> n_b$ (or $n_f=n_b$ with positive supertrace), which allows to change the trend of the potential and start increasing again as $R\rightarrow 0$. By varying the fermionic masses, we can generate this vacuum in dS, Minkowski or AdS (see Fig.\ref{fig:dS2}). The heavier the fermions are, the deeper the AdS vacuum is; crossing Minkowski when $m_f\sim \Lambda_D^{1/D}$. Let us assume now that our EFT is part of a $D$-dimensional landscape of theories that can be scanned by varying $m_f$ and that an EFT with vanishing or very small fermionic masses is part of this landscape. We can then start with $m_f\simeq 0$ and increase the masses so that the minimum is generated at smaller and smaller vacuum energy. 
 In such a case, the theory would be inconsistent with the AdS Distance Conjecture, as we can go smoothly from positive to negative minima without having an infinite tower of states coming down. Notice that we cross $V_0\rightarrow 0$ at a finite value of the radius $R\sim (\Lambda_D)^{-1/D}$, so the KK tower remains massive. There are only two possible ways out:
\begin{itemize}
\item For some reason we cannot vary the fermionic mass arbitrarily but there is an upper bound of the form $m_f \lesssim \Lambda_D^{1/D}$. This way, the scanning of $D$-dimensional vacua stops before crossing $V_0\rightarrow 0$ towards negative vacuum energy.
\item There is a correlation between the fermionic masses and $\Lambda_D$ such that $V_0\rightarrow 0$ only at $R \rightarrow \infty$. This occurs if in the original $D$-dimensional theory the cosmological constant and $m_f$ are related as $m_f \sim \Lambda_D^{\alpha_{D}}$. In general, this would imply a huge fine tuning hard to justify, unless indeed these fermions are part of the infinite tower satisfying the AdS Distance conjecture in $D$ dimensions. In such a case, we would get two towers becoming light when $V_0\rightarrow 0$; the original tower and the KK tower. It can be shown that at the minimun the KK tower which will go as $m_{KK}\sim V_0^{\frac{1}{d}}$. This singles out the particular value $\alpha_{d}= \frac{1}{d}$. The original tower will also become light. In fact it will do so with a power $\alpha_{d}$ which, unlike the KK tower, depends on $\alpha_{D}$. They can be shown to be related by $\alpha_{D}=\alpha_{d} - \frac{2 \alpha_{d} ^2+\alpha_{d} -1}{2\alpha_{d} -d^2+3}$ \cite{tom}.
\end{itemize}

We thus arrive at the following statement:

\begin{center}
{\bf Claim 3:} {\it A $D$-dim de Sitter vacuum satisfying $(-1)^{k+1} \text{Str} \, (M^{2k})>0$ for the first non-vanishing supertrace is inconsistent with quantum gravity unless there is a surplus of light fermions 1) with mass $m\lesssim \Lambda_D^{1/D}$ or 2) that are part of an infinite tower in higher dimensions scaling as $m_f\sim \Lambda_D^\alpha$.}
\end{center}

  This applies, in particular, both to non-susy theories with $n_f>n_b$ or theories with $n_f=n_b$ and spontaneously broken supersymmetry. It is interesting to check this in the Standard Model. In this case the vev of the Higgs field allows us to scan along the neutrino masses at least until we reach their experimental value. In a companion paper \cite{companion}, we study this in detail and find that the SM must have Dirac neutrinos and that the lightest neutrino must be lighter than the scale of the observed cosmological constant, realising option 1) above. Notice that the option 2) is less natural from the perspective of the SM, as it would imply that there is an infinite tower of states  starting with a mass of order the neutrino scale (or that the neutrinos are part of the tower).

So far we have conducted the discussion in terms of the total number of degrees of freedom, but the same results apply if $n_b$ and $n_f$ refer to the number of degrees of freedom below some finite energy scale. Any local (even metastable) AdS minimum generated because of a surplus of fermions will be equally inconsistent with this swampland conjecture, unless these fermions are light enough so that the minimum takes place at positive vacuum energy. However, the minimum only appears if the fermions are not closely followed by some additional bosons with a mass only slightly bigger (they can prevent the formation of a minimum if they do not leave room enough for the potential to change its behaviour). Hence, the determination of local AdS minima becomes more model-dependent.

  As a final comment, we would like to remark that we are assuming that the theory with massless (or very light) fermions is part of the landscape, so we start scanning down from positive values of the vacuum energy towards negative values. This is especially justified if supersymmetry is restored at some high energy scale or if the fermionic masses are mainly induced via a Higgs mechanism such that the symmetry restoration point is part of the landscape of consistent theories. However, if instead of scanning on the masses, one takes the perspective of scanning on different values of $\Lambda_D$ starting from $\Lambda_D<0$ towards positive values, one would conclude that the upper bound on the fermionic masses is actually a lower bound, implying in turn an upper bound on the value of the cosmological constant\footnote{We thank Eran Palti for useful comments on this regard.}.

\begin{figure}
	\centering{}
	\label{diagrama3}
	\includegraphics[scale=0.25]{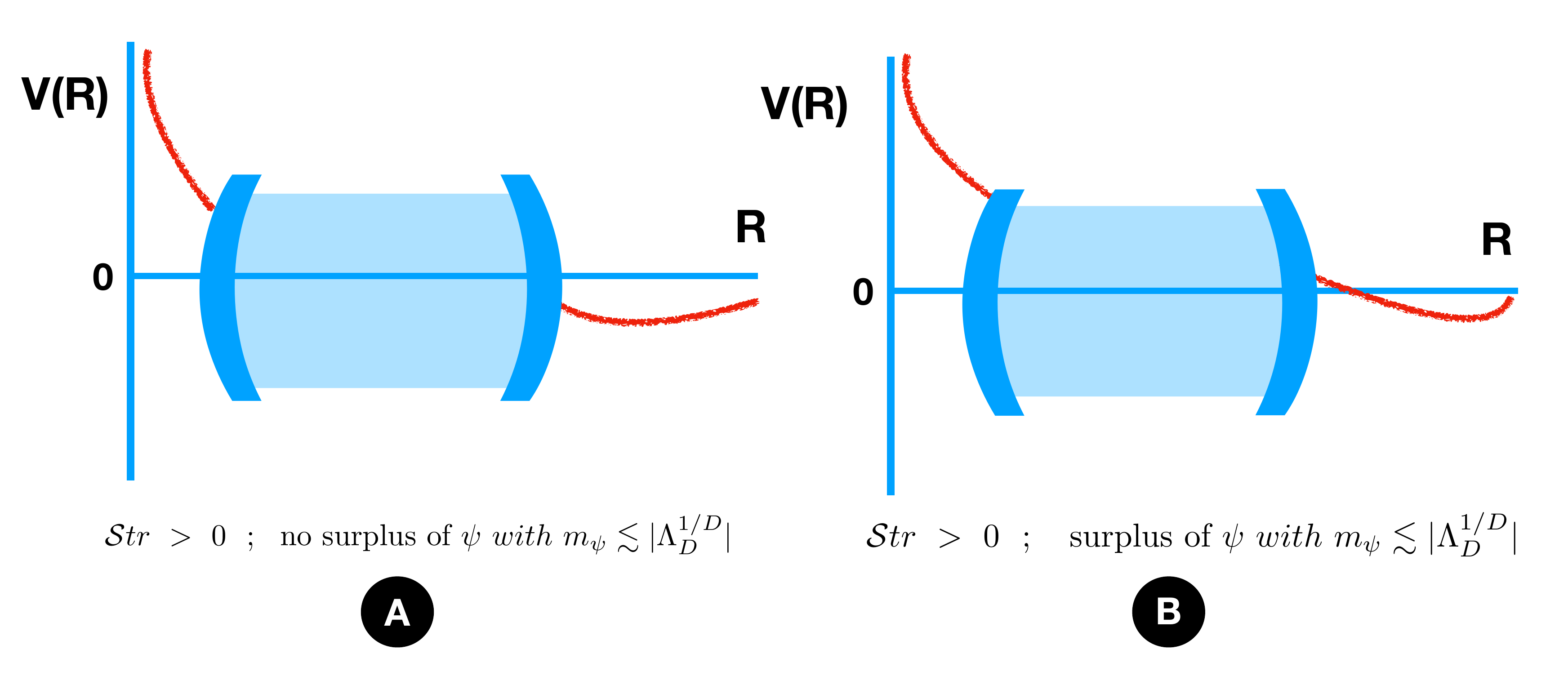} 
	\caption{\footnotesize Schematic representation of the radion $(D-1)$-dimensional potential from  $D$-dimensional  AdS vacua. The theory is in the swampland and light fermions cannot make it consistent.}
	\label{fig:AdS1}
\end{figure}

 \section{Constraints on Anti de Sitter vacua}

\subsection{From Non-SUSY AdS Conjecture}

The results are very similar to the case of Minkowski vacua. The only difference is that the potential will always approach zero from negative values as $R\rightarrow \infty$, regardless of whether the massless fields are predominantly bosons or fermions. This is due to the tree-level negative contribution from the higher dimensional cosmological constant. Hence, if the potential tends to positive values at $R\rightarrow 0$ due to having  $(-1)^k \text{Str} \, (M^{2k})>0$, a global AdS minimum will be necessarily generated, and this cannot be avoided by having additional massless fermions as in the Minkowski case (see Fig.\ref{fig:AdS1}). The results are then as follows:

\begin{center}
{\bf Claim 4:} {\it A  $D$-dim AdS vacuum satisfying $(-1)^{k+1} \text{Str} \, (M^{2k})>0$ for the first non-vanishing $k$ is
inconsistent with quantum gravity.}
\end{center}

If $\text{Str} \, (M^{0})=n_f-n_b>0$, then we get a non-susy theory inconsistent with the conjecture. If susy is broken spontaneously so this zero-th supertrace vanishes ($n_f=n_b$) but the first non-vanishing supertrace satisfies $(-1)^k \text{Str} \, (M^{2k})>0$, then the theory is also  inconsistent.

Recall that these claims hold under the assumption that a possibly higher dimensional bubble instability is not inherited by the lower $d$-dim dimensional vacuum, so that the fate of the $d$-dim vacuum can be simply determined from the potential for the radion. This implies non-trivial constraints on the radius $\rho$ of the bubble, since it must satisfy
$
\rho >V_0^{-1/4}
$
to avoid describing a bubble instability in lower dimensions. This is better justified the deeper the AdS vacuum is.

Hence, we should interpret the above claims as follows. Even if the non-susy $D$-dimensional AdS vacua are unstable, they are still inconsistent with quantum gravity unless the above condition on the spectrum is satisfied or the higher dimensional instability can be inherited by the lower dimensional vacuum. 

\subsection{From AdS Distance Conjecture}

\begin{figure}
	\label{sketchads}
	\centering{}
	\includegraphics[scale=0.25]{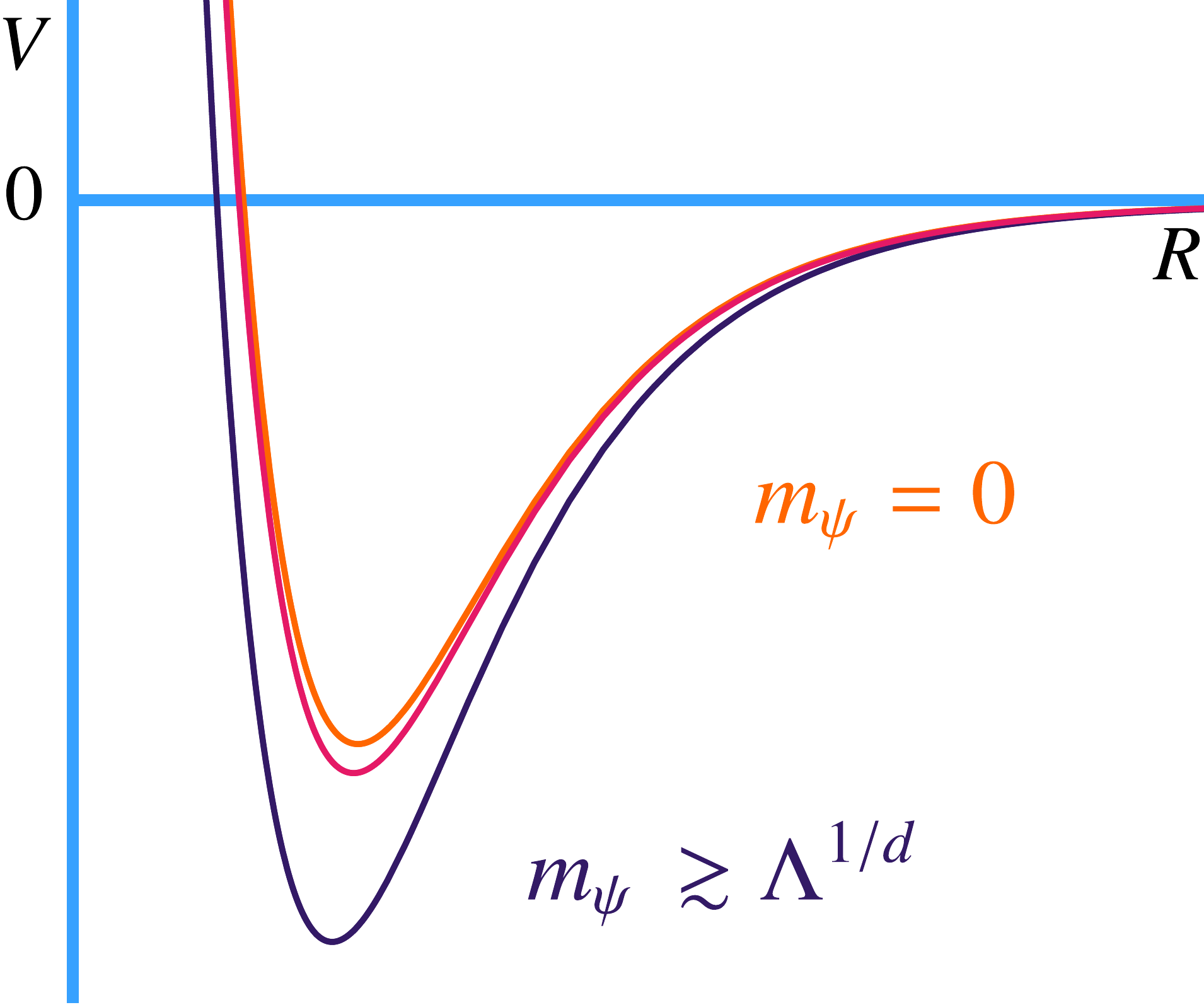} 
	\caption{\footnotesize Schematic representation of the radion $(D-1)$-dimensional potential from  $D$-dimensional  AdS vacua. We find that, as the fermions become lighter, $\Lambda_{d}$ never goes to 0, since at some point the potential is dominated by the negative $D$-dimensional cosmological constant, which, in this plot, does not change.  }
	\label{fig:AdS2}
\end{figure}

Finally, consider an AdS vacuum satisfying $(-1)^k \text{Str} \, (M^{2k})>0$ so that a lower dimensional AdS vacuum is indeed generated upon compactification. The question now is whether this vacuum is inconsistent with the AdS Distance Conjecture. For it to be inconsistent, we would need to be able to vary the masses in such a way that we can go from negative to positive minima without having a tower of states coming down. However, there is no way to do this while keeping fixed the higher dimensional $\Lambda_D$. If we decrease the masses, the minimum will take place at larger and larger values  (see Fig.\ref{fig:AdS2}) of the radius until we reach the value $R_0 \sim \Lambda_{D}^{-1/D}$ when the fermions are massless. At this point, the vacuum energy is given by
\beq
V_0\sim r^{\frac{(D-1)}{(D-3)}}\Lambda_{D}^{\frac{(D-1)(D-2)}{D(D-3)}},
\eeq
so it is non-vanishing even if the fermions are massless. Therefore, a family of AdS vacua scanned by varying the masses for fixed $\Lambda_D$ is consistent with the AdS distance conjecture, as we never cross Minkowski space this way. The only way in which $V_0$ can vanish is if we consider scannings which affect $\Lambda_D$ such that we can send $\Lambda_D\rightarrow 0$. However, in that case there will be two candidate towers getting light so that the AdS Distance Conjecture is also satisfied, as we explain in the following.

First, notice that if varying $\Lambda_D$ is a valid trajectory probing different $D$-dim EFTs, there should already be an infinite tower of states in $D$-dimensions with masses $m_{D}$ coming down as in Eq. \eqref{ADC},
since otherwise the $D$-dim vacuum itself would be inconsistent with the AdS Distance conjecture. 
Clearly, if $\alpha_D \geq 1/2$ there is no scale separation, so we can no longer use an EFT as $\Lambda_D\rightarrow 0$. However if $\alpha_D<1/2$ there is scale separation and the lighter fields of the tower will contribute to the Casimir energy in the $d$-dimensional theory in a calculable manner. The rest of the particles in the theory which are not part of the tower will not play any role if their masses do not change in the scanning. This is because at some point the lightest particles in the tower will be less massive and more numerous, so they will dominate the Casimir energy.  In the case in which the tower is made of fermions, AdS vacua will be generated. As it was recently shown in \cite{tom}, from the perspective of the lower $d$-dim AdS vacuum, the tower will scale as
\beq
m_{d}\sim \left|\Lambda_d\right|^{\alpha_d}  M_{d}^{1-2\alpha_d} \ .
\eeq
with $\alpha_{D}$ and $\alpha_{d}$  related by $\alpha_{D}=\alpha_{d} - \frac{2 \alpha_{d} ^2+\alpha_{d} -1}{2\alpha_{d} -d^2+3}$. In fact, one can check that the above relation implies $\alpha_d<\alpha_D$, so the tower becomes light at a slower rate from the lower dimensional perspective. This can make hard to satisfy the strong version of the AdS Distance conjecture, i.e. $\alpha\geq \frac12$, upon compactification. The value $\alpha_D=1/D$ was also highlighted in  \cite{tom} as a special value which is robust under dimensional reduction, since it implies in turn $\alpha_d=1/d$.
  
Secondly, we would like to remark that the value of the radius at the minimum diverges $R_0\rightarrow \infty$ as  $V_{0}\sim \Lambda_{d}\rightarrow 0$. Hence, regardless of the previous $D$-dimensional tower, we will also have the KK tower becoming light as
\beq
M_{KK}\sim \Lambda_{d}^\frac{1}{d}
\eeq
in Planck units. This is the same result for the exponent that we found for the cases of $dS_D$ and $M_D$. Again, this is consistent with the mild version of the AdS Distance Conjecture, but not with its stronger version requiring $\alpha\geq \frac12$.

To conclude, AdS vacua seem to be consistent with the mild (but not the strong) version of the AdS Distance conjecture upon circle compactification.  In principle, it may also be possible to engineer more cumbersome situations which could be inconsistent with the conjecture by generating several local maxima and minima at intermediate values of the radius. In such a case, there might be some scanning trajectory varying the masses in a non-homogeneous way  that forces $V_0$ to vanish at finite radius. However, this becomes highly model-dependent and it is hard to extract any general statement.

\begin{table*}[t]\begin{center}
		
		\renewcommand{\arraystretch}{1.00}
		\begin{tabular}{|c||c|c|}
			\hline
			Vacua &  non-SUSY\ AdS&  AdS\ distance \\
			\hline \hline
			$M_D$ &   violated\ (unless \ surplus &  \ $\alpha=1/d$ \\
			&   of\  massless\ fermions)  &     \\
			\hline
			$dS$ &  violated \ (unless \ surplus&  violated$^*$\ (unless \ surplus \\
			&of\  fermions\ $m_f\lesssim \Lambda^{1/D}$) &  fermions\ $m_f\lesssim \Lambda^{1/D}$)   \\
			\hline
			$AdS$  & violated &        \\
			\hline
		\end{tabular}
	\end{center}
	\caption{Summary of the constraints for different types of $D$-dimensional vacua. The constraints apply whenever the first non-vanishing
		supertrace is positive, $(-1)^{k+1}\text{Str} \,M^{2k}>0$. The asterisk indicates that there is no violation if the light fermions are part of a distance conjecture tower.}
\end{table*}

\vspace{-0.3cm}
 
 \section{ A light fermion Swampland conjecture}

 The necessity of light-fermions in the theory has appeared often in the previous sections, in particular in the case of Minkowski and dS vacua.
 One is tempted to conjecture that the presence of these light fermions may be a  general feature of these vacua, beyond the motivation
 we have found from AdS swampland conjectures.
 In particular the constraints from the non-SUSY AdS conjecture  on $M_D$ and $dS_D$ vacua may be summarized in the following

\vspace{0.3cm}

{\textbf{ Light fermion Swampland conjecture:} } 

\vspace{0.1cm}

{\it In a SUSY-broken theory with $\Lambda_D\geq 0$ and  positive  first non-vanishing supertrace
 $(-1)^{k+1}{\cal S}trM^{2k}$, which is  consistent with quantum gravity,
there must exist a surplus of light fermions with masses
$m\lesssim \Lambda_D^{1/D}$}.

\vspace{0.3cm}

Interestingly, the need for light fermions  with $m\lesssim \Lambda_D^{1/D}$ in dS  is also independently supported by the  AdS distance conjecture,
as we saw in section IV. This may point to a more general principle underlying this light fermion conjecture.
Note also that this conjecture directly relates UV data (from the supertraces)  with IR data (massless/light fermions).
 It would be interesting to investigate whether something like this  light fermion swampland conjecture might also hold in AdS \footnote{It seems that light fermions are also a common feature in AdS string theory vacua. We thank Juan Maldacena and Tom Rudelius for comments on this regard.}, even if it is not supported by the AdS Swampland conjectures.

It would be interesting to look for string theory tests of this light fermion conjecture. 
In order to do that we would need 
to study examples of  non-SUSY $dS_D$ and $M_D$ string vacua, which are not very abundant. An example which comes close to be a test 
is the non-SUSY 10D heterotic string with gauge group $SO(16)\times SO(16)$, which has no tachyons
 \cite{AlvarezGaume,Dixon}. This seems to be a consequence of the property of ``missaligned supersymmetry'' which has been shown to be
 present in a number of non-SUSY string vacua  \cite{Dienes0}.
  The supertraces over all the string spectrum of this 10D model  have been computed  \cite{Dienes,Dienes2,Abel,Wrase} and have been shown to vanish up to order  $M^8$, with 
  $M$ the string mass.
   The contribution at order $M^8$ to  $(-1)^{1+k}{\cal S}trM^{2k}$  turns out to be positive for $k=4$. 
  Thus, according to the above conjecture, there should be a surplus of massless fermions in the theory.
    Indeed, the massless spectrum of $SO(16)\times SO(16)$ contains the 10D Gravitino, Kalb-Ramond and dilaton bosonic fields,
  without SUSY partners, the gauge bosons of the group and spinors in the representations $(128,1)+(1,128)$ and $(16,16)$.  Altogether the reader may check that 
  at the massless level there is a surplus of 2112 fermions.  Although the traces computed over the full string spectrum require careful regularization \cite{Seiberg}
  and hence are not just given by our simple expression for the supertraces above, the physical principle is essentially the same. 
Notice that this example has a runaway potential at one-loop and hence it is not a  $dS$ vacuum. %$properly speaking, and the motivation for our conjecture coming from the AdS swampland conjectures does not directly apply to this case. 
Yet, it is very tantalising that it shows a connection between a positive supertrace and the existence of 
  a surpluss of massless fermions. This may indicate that the conjecture still applies to runaway dS potentials, requiring in this case a surplus 
  of (strictly massless) fermions.
    
  %There is another non-SUSY  tachyon-free $10D$ theory with gauge group $Sp(32)$\cite{Sugimoto,Sagnotti}. It is obtained by locating  anti-D9 branes on top of
 % $O(9)^+$ orientifolds in Type I string theory.  Such orientifolds are exotic in that they have positive tension.
   %  In  the massless sector this theory has a SUSY 10D supergravity multiplet, the gauge group and  496 fermions in the antisymmetric representation.
  % The reader may check that there is a surplus of 8  bosonic massless states in this case. This however is not in contradiction with the
 %  conjecture since in this case the vacuum energy has the form  $V\sim [1\ -\ (g_sI_p/(2\pi)e^\phi]$ and the theory is driven to a non-perturbative 
   %regime \cite{Wrase,Sagnotti}, which would possibly hint a negative supertrace. It would be interesting to study this in more detail.

From the phenomenological point of view this conjecture may have profound implications. The universe seems to be in a dS era and in the 
the SM the first non-vanishing supertace ($k=0$) is positive, since it contains more fermions than bosons. As we said, this would imply the necessity of 
fermions lighter than the observed c.c. scale $\Lambda_4^{1/4}\simeq 10^{-2}\ eV$.  Interestingly, the SM can provide for such fermions if the lightest neutrino 
has a Dirac mass term and is sufficiently light. As we said this has been analyzed in detail in  \cite{IMV1,HS,2toro}. This would also apply to SUSY versions of the SM if the  $M^2$ supertrace 
is positive as happens in many models like e.g. Split SUSY \cite{Nima,Gian}.  It has also been pointed out  in \cite{IMV2,Gonzalo} that this can provide for an understanding of the hierarchy 
problem of the SM.

\section{Final comments and conclusions}

In this paper we have explored the constraints on D-dimensional vacua arising from requiring that circle compactifications of such a theory are consistent with the AdS swampland conjectures. This puts constraints on the physical spectra of the D-dimensional theory that guarantee that the Casimir energy potential does not generate $(D-1)$-dimensional AdS vacua that would violate the swampland conjectures and, therefore, be inconsistent with quantum gravity. The constraints obtained for the different cases are summarized in Table 1.

In the case of starting with Minkowski or de Sitter vacua, it seems that theories with the first non-vanishing supertrace  Eq. \eqref{supertrace_defined} satisfying $(-1)^{k+1}{\cal S}trM^{2k}$are inconsistent with the Non-SUSY AdS conjecture unless the number of light fermionic degrees of freedom with masses $m\lesssim \Lambda^{1/D}$ exceeds the bosonic one. This rules out, for example, non-susy theories with more fermions than bosons, or theories with susy spontaneously broken such that the bosons are more massive than the fermions (like e.g.  in Split SUSY); unless there is the aforementioned surplus of light fermions (i.e. neutrinos in the SM). The same results are obtained from applying the AdS Distance conjecture to  dS vacua, which provides additional support that it is independent of the stability of the vacua. More concretely, consistency with this conjecture in dS implies either the presence of a surplus of light fermions or that the lightest fermions of the theory are part of an emerging  tower of particles.

This motivates us to formulate a {\it light fermion swampland conjecture}, which requires the existence of fermions with masses $m\lesssim \Lambda^{1/D}$ in a vacuum with $\Lambda\geq 0$ and the above condition on the supertrace. Interestingly, light fermions seem to be a common feature in string theory but also in our world, where neutrino masses are  of order the cosmological constant scale. According to our work, this numerical coincidence could have a quantum gravity origin, which could bring a new perspective into the EW hierarchy problem, as we first pointed out in  \cite{IMV1,IMV2} (see also \cite{2toro,Gonzalo,companion,tom}).

In the case of  Minkowski vacua, the AdS distance conjecture is automatically satisfied by the KK tower in the absence of massless fermions, so
no additional support for massless fermions is obtained. An interesting fact, though, is that  the KK tower scales as $M_{KK}=c\Lambda_d^{\alpha_d}$
with $\alpha_d=1/d$ in the lower $d$-dimensional theory. This is in conflict with the strong version of the AdS distance conjecture which states that 
$\alpha \geq 1/2$ for non-SUSY vacua.
 The same scaling of the KK tower occurs in circle compactifications of dS and AdS vacua when taking the limit $\Lambda_D\rightarrow 0$. Hence, again the mild but not the strong version of the AdS distance conjecture is satisfied.
This universal scaling of the KK tower both in compactifications of Minkowski, dS and AdS vacua suggests that the value  $\alpha_{d}=\frac{1}{d}$ should be consistent with quantum gravity. This could  suggest a modification of the Strong AdS Distance conjecture to $\alpha_{d} \geq \frac{1}{d}$, thus allowing a certain level of scale separation. In particular, it would imply  that scale-separated 4D AdS vacuum in  \cite{DGKT,CFI}  with $\alpha=7/18$ would be consistent with this modification of the conjecture. Interestingly, the same scaling $\alpha_{d}=\frac{1}{d}$ was recently highlighted in \cite{tom} from a different perspective. Another possible interpretation of our results is that the Strong AdS Distance conjecture with $\alpha_{d} \geq \frac{1}{2}$ indeed holds and therefore all the above circle compactifications of AdS and Minkowski satisfying the above condition on the supertrace are somehow inconsistent with quantum gravity. In fact, the Non-SUSY AdS conjecture also suggests that such AdS vacua are inconsistent, but this comes with additionals assumption regarding the stability of the vacuum. Notice that, unlike what happens in $M_D$ and $dS_D$, light fermions are not enough to cure the $AdS_D$ vacuum.

For future investigation, it would be interesting to search for additional string theory evidence for our light fermionic conjecture. Moreover, it would be interesting to look for some bottom-up physics rationale for such a correlation between the mass of the lightest fermions and the cosmological constant. Currently, understanding the naturalness issues and hierarchy problems observed in our universe is one of the most challenging questions in High Energy Physics, and quantum gravity consistency may be the missing piece in this puzzle.

\vspace{-0.2cm}
\begin{acknowledgments}

 We thank Keith Dienes, Alvaro Herr\'aez, Fernando Marchesano, Miguel Montero, Tom Rudelius, A. Uranga  and Timm Wrase for useful discussions. 
 This work is  is supported  by  the  Spanish  Research  Agency  (Agencia  Espa\~nola  de  Investigaci\'on) through  the  grants  IFT  Centro  de  Excelencia  Severo  Ochoa  SEV-2016-0597, the grant GC2018-095976-B-C21 from MCIU/AEI/FEDER, UE and the grant PA2016-78645-P. 
   E.G. is supported by the Spanish FPU Grant No. FPU16/03985. The research of IV was supported by a
grant from the Simons Foundation (602883, CV).
\end{acknowledgments}

%\newpage

\end{document}